# Force Method in a Pseudo-potential Lattice Boltzmann Model

Anjie Hu[1*], Longjian Li[1], Rizwan Uddin[2]

a. Key Laboratory of Low-grade Energy Utilization Technologies and Systems of Ministry of Education, Chongqing 400030, PR China

b. Department of Nuclear, Plasma and Radiological Engineering, University of Illinois at Urbana Champaign, Urbana, IL 61801, USA

**Abstract**

Single component pseudo-potential lattice Boltzmann models have been widely studied due to their simplicity and stability in multiphase simulations. While numerous model have been proposed, comparative analysis and advantages and disadvantages of different force schemes is often lacking. A pseudo-potential model to simulate large density ratios proposed by Kupershtokh et al. [1] is analyzed in detail in this work. Several common used force schemes are utilized and results compared. Based on the numerical results, the relatively most accurate force scheme proposed by Guo et al. [2] is selected and applied to improve the accuracy of Kupershtokh et al.'s model. Results obtained using the modified Kupershtokh et al.'s model [1] for different value of $\tau$ are compared with those obtained using Li et al. [3] 's model. Effect of relaxation time $\tau$ on the accuracy of the results is reported. Moreover, it is noted that the error in the density ratio predicted by the model is directly correlated with the magnitude of the spurious velocities on (curved) interfaces. Simulation results show that, the accuracy of Kupershtokh et al.'s model can be improved with Guo et al. [2] 's force scheme. However, the errors and $\tau$ 's effects are still noticeable when density ratios are large. To improve the accuracy of the pseudo-potential model and to reduce the effects of $\tau$, two possible methods were discussed in the present work . Both, a rescaling of the equation of state and multi-relaxation time, are applied and are shown to improve the prediction accuracy of the density ratios.

---

[*] Corresponding author. Tel.: +86 15310623987
 E-mail address: anjie@cqu.edu.cn (A. Hu)

**1. Introduction**

Lattice Boltzmann equation (LBE) method [4], also known as Lattice Boltzmann method (LBM), has attracted a significant attention due to its potential to solve problems at the mesoscopic scale. From its origins in lattice gas automation method, it has been developed into a numerical method to simulate fluid flows and other nonlinear problems. One promising application of LBE method is multiphase flow simulation. Due to its kinetic nature, a well-developed theoretical basis, and the ability of self-capturing the interface, LBE method has many advantages when simulating multiphase problems. Several LBE multiphase models have thus been developed. These models can be summarized into four categories: color models [5], pseudo-potential methods [6], free energy models [7, 8] and kinetic models [9-11]. Gunstensen et al. [5] proposed the first color LBE model by labeling components and particles by colors in the LBE model. Several extensions were developed based on Gunstensen et al.'s model and have been successfully applied to complex interfacial flows [12, 13]. However, these models suffer from several limitations, such as the anisotropy of surface tension and spurious currents [14]. Due to their simplicity and stability at a high density ratio, pseudo-potential models first proposed by Shan and Chen [6] are widely used, but they also have drawbacks such as spurious currents [14]. The first free energy type LBE model was proposed by Swift et al. [7]. However, it is restricted to low density ratios, and the early free energy LBE models often suffered from Galilean invariance [8]. Kinetic LBE models, as the name suggests, are based on kinetic methods. A typical kinetic LBE model is by He Shan Doolen [9], which is based on a modified Boltzmann equation. With a special discrete method, Lee and Lin [15] successfully extended the He-Shan-Doolen model for large density ratios. However, in addition to the sensitivity of the discrete approach, it has been shown that the mass conservation of these models is weak for large density ratios [16].

These LBE multiphase models have been widely used in simulations. However, most of the above models are limited to multiphase flows with small density ratios. To solve this problem, several additional LBE multiphase models for large density ratios were proposed [15, 17-19]. Among these models, single component pseudo-potential models show promise to solve large density ratio flows since they are stable for large density ratios without fancy numerical methods. However, Yuan and Schaefer [19] found that the stabilities of these models vary with equations of state introduced in the pseudo-potential models. To address this issue, they developed the large density ratio pseudo-potential model by choosing an appropriate equation of state (EOS). However, it has been shown that the pseudo-potential models are consistent with thermodynamic theories only when the EOS takes a special exponent form [20]. The stability of the pseudo-potential is related to the pressure tensor which varies with the inter-particle interaction force models and the LBE force schemes adopted in the model [3, 21].

To address these problems, several approaches have been proposed to reduce the thermodynamic error and to increase the stability of the pseudo-potential method. The most common approach is the multi-range pseudo-potential model, developed by Sbragaglia et al. [22], which combines the nearest-neighbor interactions and the next-nearest-neighbor interactions. Though much improved, the introduction of the next-nearest-neighbor interactions leads to difficulties in programming especially for the boundary conditions. Li et al. [3] recently put forward a method to reduce the thermodynamic error by introducing an additional term in the force scheme. It successfully improved the stability without adding much numerical cost. However, the special treatment of the inter-particle interaction force is developed specifically for the force scheme proposed by Guo et al. [2].

In a parallel effort to reduce the thermodynamic error, Kupershtokh et al. [1, 23, 24] pointed out that the scale of the EOS is the main reason for the stability of the pseudo-potential model. They also developed an interparticle-force model by combining two nearest-neighbor interactions models and adjusting the scale of the reduced EOS. Later, Hu et al. [25] extended this method to general EOS.

The development of Kupershtokh et al.'s model is however somewhat adhoc. Thus the choice of the parameters introduced in the model lacks theoretical foundation. Moreover, some studies have shown that the exact difference method (EDM) force scheme [26] applied in Kupershtokh et al.'s work leads to error terms in the corresponding macroscopic equation, and thus the numerical problem being solved is different from the original macroscopic problem [3, 21]. Huang et al. did attempt to integrate different LBE approaches, and provided some theoretical foundation for the Kupershtokh et al. model. However in Huang et al.'s work [21], the density distributions of the EDM force scheme vary with the relaxation time, which is not the case in Kupershtokh et al.'s work.

We here report a numerical error analysis of the Kupershtokh et al.'s model for EDM force scheme. We then extend and improve the model by applying the force scheme proposed by Guo et al. [2], instead of using the EDM forcing scheme, thus eliminating the error in the corresponding macroscopic equation. Finally, numerical results obtained using the improved method developed here and those obtained using Li et al.'s method [3], which adopted a different approach to approximately satisfy the thermodynamic constraints, are compared.

Rest of the paper is organized as follows. The pseudo-potential LB model is briefly introduced in Section 2. In Sec. 3, interparticle interaction force calculation methods and the forcing schemes are theoretically analyzed. Numerical investigations and comparisons are presented in Sec. 4. Finally, conclusion are drawn in Sec. 5.

## 2. Pseudo-potential model

In the LBE method, the motion of the fluid is described by evolution of the density distribution function. The evolution equation can be written in the form of the BGK operator [27] as

$$f_\alpha(\boldsymbol{x}+\boldsymbol{e}_\alpha\Delta t, t+\Delta t) - f_\alpha(\boldsymbol{x},t) = -(f_\alpha(\boldsymbol{x},t) - f_\alpha^{eq}(\boldsymbol{x},t))/\tau + \boldsymbol{F}_\alpha, \qquad (1)$$

where $\tau$ is the reduced relaxation time, $f_\alpha(\mathbf{x},t)$ is the density distribution function of particles at node $\mathbf{x}$ and time $t$, and $\mathbf{e}_\alpha$ is the velocity where $\alpha = 0, 1, 2 \cdots N$. The right side of the equation is a collision operator, $\mathbf{F}_\alpha$ is the force term, $f_\alpha^{eq}(\mathbf{x},t)$ is the equilibrium distribution function which can be represented in the following form for the two-dimensional nine-velocity (D2Q9) lattice:

$$f_\alpha^{eq}(\mathbf{x},t) = w_\alpha \rho(\mathbf{x},t)[1+(e_\alpha \cdot \mathbf{u}^{eq})/c_s^2 + (e_\alpha \cdot \mathbf{u}^{eq})^2/(2c_s^4) - (\mathbf{u}^{eq})^2/(2c_s^2)], \quad (2)$$

where $w_0 = 4/9$, $w_\alpha = 1/9$ for $\alpha = 1, \cdots, 4$, and $w_\alpha = 1/36$, for $\alpha = 5, \cdots, 8$. Sound speed $c_s$ is $c/\sqrt{3}$, where $c = \delta_x/\delta_t$, and $\delta_x$ and $\delta_t$ are lattice spacing and time step respectively (both of them are equal to 1 in the following simulations). $\mathbf{u}^{eq}$ is the equilibrium velocity which depends on the force schemes. $\rho(\mathbf{x},t)$ is density, given by

$$\rho(\mathbf{x},t) = \sum_{\alpha=0}^{8} f_\alpha(\mathbf{x},t) = \sum_{\alpha=0}^{8} f_\alpha^{eq}(\mathbf{x},t). \quad (3)$$

In single component pseudo-potential model, the phase separation is achieved by introducing an interparticle interaction force between particles at neighboring lattice sites. For interaction between only the nearest neighbors, there are two commonly used interparticle interaction force models. First of these is the effective density type, proposed by Shan and Chen [6, 20], which can be written as

$$\mathbf{F}_1(\mathbf{x},t) = -G\psi(\mathbf{x},t)\sum_\alpha w(|e_\alpha|^2)\psi(\mathbf{x}+e_\alpha \delta_t,t), \quad (4)$$

where $G$ is the interaction strength, $w(|e_\alpha|^2)$ are the weights, and $\psi(\mathbf{x},t)$ is effective density. For the case of nearest neighbor interactions on the D2Q9 lattice, the weights $w(|e_\alpha|^2)$ are $w(1) = 1/3$ and $w(2) = 1/12$.

The second one is the potential function model proposed by Zhang et al. [28],

which can be written as

$$F_2(x,t) = -\sum_\alpha w(|e_\alpha|^2) U(x+e_\alpha \delta_t, t),  \quad (5)$$

where $U(x,t)$ is the potential function which is equal to $G\psi^2(x,t)/2$.

To improve the stability of the pseudo-potential model, Kupershtokh et al. [1] proposed a hybrid model obtained by combining these two models mentioned above, which is given by

$$F_3(x,t) = -A\sum_{\alpha=0}^{8} \omega_\alpha U(x+e_\alpha) e_\alpha - (1-A) G\psi(x,t) \sum_{\alpha=0}^{8} \omega_\alpha \psi(x+e_\alpha, t) e_\alpha. \quad (6)$$

In practice, both, effective density and potential function, models can be obtained by introducing a non-ideal EOS [19]:

$$p_0 = c_s^2 \rho + cG[\psi(\rho)]^2 / 2 = c_s^2 \rho + U(\rho). \quad (7)$$

By choosing this form of the EOS, the interaction strength $G$ gets canceled in Eqs. (4) and (6) [19], and hence the interaction force is depend only on the EOS.

## 3. Theoretical analysis

Kupershtokh et al.'s force model shows a great improvement compared with the original Shan and Chen's model. However, the theoretic analysis have rarely been mentioned in early literatures, and it has been shown that the EDM force scheme applied in the model leads to extra terms when compared with Navies-Stokes equation. To fill the gap, the detailed analysis of the force model will be made in Sec. 3.1, and the EDM will be theoretically compared with two other common used force schemes in Sec. 3.2.

**3.1 Mechanical solution of Kupershtokh et al.'s force model**

To obtain the macroscopic expression corresponding to each force model, Shan's method [29] is applied in the present work. Through Taylor expansion, the leading term of the interaction force in the effective density model represented by Eq. (4) can be written as

$$F_1 = -Gc^2\left[\psi\nabla\psi + \frac{e_2 c^2}{6}\psi\nabla(\nabla^2\psi) + \cdots\right]. \tag{8}$$

Similarly, the macroscopic expressions for the potential function model (Eq. (5)) and the hybrid model (Eq. (6)) can be written as:

$$F_2 = -\frac{1}{2}Gc^2\left[2\psi\nabla\psi + \frac{e_2 c^2}{3}\psi\nabla(\nabla^2\psi) + 3\frac{e_2 c^2}{3}(\nabla^2\psi)\nabla\psi + \cdots\right]. \tag{9}$$

$$F_3 = -Gc^2\left[\psi\nabla\psi + \frac{e_2 c^2}{6}\psi\nabla(\nabla^2\psi) + \frac{Ae_2 c^2}{2}(\nabla^2\psi)\nabla\psi + \cdots\right]. \tag{10}$$

The corresponding pressure tensors can be obtained from the force expressions. However, the pressure obtained by integrating the Taylor expanded force form may be inconsistent with the pressure tensor obtained by the Chapman-Enskog expansion. To overcome this problem, Shan [29] pointed out that the pressure tensor should be derived from the volume integral of the original force expression.

According to Shan [29], the corresponding pressure tensors for the three different force models can be written as:

$$P_1 = -\frac{1}{2}G\psi(x)\sum_{\alpha=1}^{N} w(|e_\alpha|^2)\psi(x+e_\alpha)e_\alpha e_\alpha, \tag{11}$$

$$P_2 = -\frac{1}{4}G\sum_{\alpha=1}^{N} w(|e_\alpha|^2)\psi^2(x+e_\alpha)e_\alpha e_\alpha, \tag{12}$$

$$P_3 = AP_2 + (1-A)P_1. \tag{13}$$

Expanded in a Tayler series, the pressure tensors can be rewritten as:

$$\boldsymbol{P}_1 = \left( \rho c_s^2 + \frac{Gc^2}{2}\psi^2 + \frac{Gc^4}{12}\psi\nabla^2\psi \right)\boldsymbol{I} + \frac{Gc^4}{6}\psi\nabla\nabla\psi + O(\partial^4\psi), \tag{14}$$

$$\boldsymbol{P}_2 = \left( \rho c_s^2 + \frac{Gc^2}{2}\psi^2 + \frac{Gc^4}{24}\nabla^2\psi^2 \right)\boldsymbol{I} + \frac{Gc^4}{12}\nabla\nabla\psi^2 + O(\partial^4\psi^2), \tag{15}$$

$$\boldsymbol{P}_3 = \left( \rho c_s^2 + \frac{Gc^2}{2}\psi^2 + \frac{Gc^4}{24}\nabla^2\psi^2 \right)\boldsymbol{I} + \frac{Gc^4}{6}\psi\nabla\nabla\psi + A\frac{Gc^4}{6}\nabla\psi\nabla\psi + O(\partial^4\psi) + O(\partial^4\psi^2)$$

$$. \tag{16}$$

Restricting the presentation here to one-dimensional two-phase equilibrium, the pressure tensor can in general be written as

$$P_c = \rho c_s^2 + \frac{Gc^2}{2}\psi^2 + \frac{Gc^4}{12}\left[ a\left(\frac{d\psi}{dn}\right)^2 + b\psi\frac{d^2\psi}{dn^2} \right] + O(\partial^4), \tag{17}$$

where $n$ represents the normal direction of the interface between the phases. Parameters $a$ and $b$ are given in Table 1 for the three force models.

Table 1. Parameters $a$ and $b$ for different models.

| Force model | a | b |
|---|---|---|
| Effective density (Eq.(4)) | 0 | 3 |
| Potential function (Eq.(5)) | 3 | 3 |
| Kupershtokh et al.'s model (Eq.(6)) | 3A | 3 |

The densities in the gas and liquid phases should satisfy the following relation [29]:

$$\int_{\rho_g}^{\rho_l} \left( p_c - \rho c_s^2 - \frac{Gc^2}{2}\psi^2 \right) \frac{\psi'}{\psi^{1+\varepsilon}} d\rho = 0, \tag{18}$$

where subscript $l$ and $g$ represent liquid and gas, respectively, $p_c = p_0(\rho_l) = p_0(\rho_g)$, and $\varepsilon = -2a/b$. Eq. (18) is called the mechanical stability condition [21]. It can be seen that although different force calculation methods are applied, the mechanical stability condition of Kupershtokh et al.'s model, which is represented by Eq. (18), is identical with Eq. (26) in Shan [29]'s work and Eq. (26) in Li et al.'s model [3]. Moreover, with the non-ideal EOS (Eq. (7)), the requirement to satisfy Maxwell construction can be written as

$$\int_{n_g}^{n_l} (p_c - p_0) \frac{1}{\rho^2} d\rho = 0. \qquad (19)$$

There are two ways to satisfy Eq. (19). One is by choosing a special effective density form which makes Eq. (18) identical to Eq. (19), as shown by Sbragaglia and Shan [30], which means only the specific EOS can be applied in the pseudo-potential model. In the second, more practical approach, in order to apply general EOS in the model, the parameter $\varepsilon$ in Eq. (18) is adjusted, as in Kupershtokh et al. [1] and Li et al. [3], to approximately satisfy the results of the Maxwell construction [3].

### 3.2 Force schemes

There are three force schemes commonly used in the pseudo-potential model: the Shan-Chen (SC) type force scheme [6] which incorporates the force by shifting the velocity in the equilibrium distribution; the Exact-Difference-Method (EDM) proposed by Kuprestokh et al. [26]; and the Guo et al.'s force scheme [2]. For SC force scheme, the evolution equation can then be written as

$$f_\alpha(\bm{x} + \bm{e}_\alpha \Delta t, t + \Delta t) - f_\alpha(\bm{x}, t) = -(f_\alpha(\bm{x}, t) - f_\alpha^{\text{eq}}(\bm{x}, t))/\tau. \qquad (20)$$

The equilibrium velocity $\bm{u}^{\text{eq}}$ is given by

$$\bm{u}^{\text{eq}} = \bm{u} + \frac{\tau \delta t \bm{F}}{\rho}, \qquad (21)$$

where $u = \frac{\sum_\alpha f_\alpha e_\alpha}{\rho}$. The actual fluid velocity is defined as $v = u + \frac{\delta t F}{2\rho}$.

Kupershtokh et al. [26] noted that the density distribution for the pseudo-potential model obtained with Shan-Chen's force scheme varied with the relaxation time $\tau$. To avoid this dependence, they proposed the so called Exact-Difference-Method (EDM) scheme [26]. In this scheme, the force term in Eq. (1) can be written as

$$F_\alpha = f_\alpha^{eq}(\rho, u + \Delta u) - f_\alpha^{eq}(\rho, u), \qquad (22)$$

where $u = \frac{\sum_\alpha f_\alpha e_\alpha}{\rho}$, and $\Delta u = \frac{\delta t F}{\rho}$. The actual fluid velocity is still defined as $v = u + \frac{\delta t F}{2\rho}$.

However, it has been shown that both SC force scheme (Eq.21) and EDM force scheme (Eq.22) lead to error terms in the corresponding macroscopic equations, and lead to the coexistence curves that are different from the mechanical solutions [3, 21]. Reason behind this discrepancy is the low precision of these two force schemes (SC and EDM). Hence, higher precision force models have been suggested as the solution to the problem. As mentioned earlier, Guo et al.'s scheme is a higher precision force scheme. It is used here to develop improved schemes for the LBE method for high density ratio problems. Guo et al.'s force model [2] is given by

$$F_\alpha = \omega_\alpha \left(1 - \frac{1}{2\tau}\right) \left[ \frac{F_i e_{\alpha i}}{c_s^2} + \frac{(F_i v_j + v_j F_i)(e_{\alpha i} e_{\alpha j} - c_s^2 \delta_{ij})}{2 c_s^4} \right] \qquad (23)$$

Here, the velocity used in the equilibrium distribution function should be equal to the actual fluid velocity, which is $v = u + \frac{\delta t F}{2\rho}$.

The macroscopic equations for the three schemes (SC, EDM and Guo's force scheme) can be obtained through the Chapman-Enskog analysis [31]. The resulting continuity and momentum equations for all three cases are given by

$$\partial_t \rho + \partial_i (\rho v_i) = 0, \tag{24}$$

$$\partial_t \rho v_i + \partial_i (\rho v_i v_j) = -\partial_i p + \partial_i (\rho \upsilon S_{ij}) + F_j + R_v, \tag{25}$$

where $v$ is the actual velocity of the fluid, $S_{ij} = (\partial_j v_i + \partial_i v_j)/2$, and $\upsilon$ is the viscosity coefficient which is equal to $c_s^2 \left(\tau - \frac{1}{2}\right)\delta_t$. Note that the continuity equations for all three cases correspond exactly to the continuity representation of the conservation of mass. However, the momentum equation has an extra term ($R_v$) when compared with the Navies-Stokes equation. For the LBE approach to exactly satisfy the momentum balance, $R_v$ should be equal to zero, and that is why it is here termed as the error term. This term for the three force schemes are

$$R_{v, sc} = -\varepsilon^2 \delta_t \left(\tau - \frac{1}{2}\right)\frac{\partial F_j}{\partial t_1} + \delta_t \partial_i \left[\upsilon\left(\tau - \frac{1}{2}\right)(\partial_i F_j + \partial_j F_i) + \delta_t \left(\tau - \frac{1}{2}\right)^2 \frac{F_i F_j}{\rho}\right], \tag{26}$$

$$R_{v, EDM} = -\varepsilon^2 \delta_t \left(\delta_t - \frac{1}{2}\right)\frac{\partial F_j}{\partial t_1} + \delta_t \partial_i \left[\upsilon\left(\delta_t - \frac{1}{2}\right)(\partial_i F_j + \partial_j F_i) + \delta_t \left(\delta_t - \frac{1}{2}\right)^2 \frac{F_i F_j}{\rho}\right],$$

$$\tag{27}$$

$$R_{v, Guo} = 0. \tag{28}$$

Both SC and EDM schemes are therefore will lead to numerical results that are not expected to match the solution of the Navies-stokes equations. However, these error terms may make the model more stable in some cases [3].

Moreover, it is not clear what role, if any, $\tau$ plays in these models. For example, model with EDM force scheme is claimed to be independent of $\tau$ [26], but Huang et al. [21] found that the density distribution found using the EDM force scheme varied with $\tau$ for large density ratio problems.

To assess the effect of $\tau$ on different force scheme models, we examine the momentum equation (Eq. 25) along with the expressions for $R_v$ (Eqs. (26-28)) Obviously, all momentum equations depend on $\tau$ because of the viscosity terms: $\partial_i(\rho \upsilon S_{ij})$. However, these terms do not influence the parameters of two-phase flows when the gradients of velocities are zero in the interfacial area. However, the existence of spurious velocities in the pseudo-potential model may lead results that vary with $\tau$. To further analyze the influence of this term, we simulated two phases with straight and curved interfaces. Velocity distributions for these two cases shown in Fig. 1. It can be seen that for the straight interface case, the spurious velocities are normal to the interface and their magnitude do not change along the interface, Hence $\partial_i(\rho \upsilon S_{ij})$ is zero in this case, which means $\tau$ does not influence density distribution. However, for curved interface, the spurious velocities along the interface are irregular, possibly leading to a non-zero contribution from the $\partial_i(\rho \upsilon S_{ij})$ term and hence dependence on $\tau$.

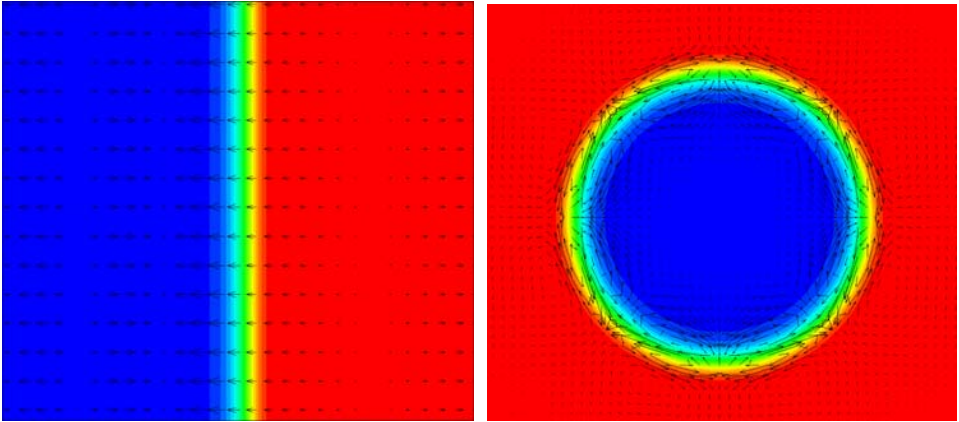

Fig. 1 Spurious velocities of straight and curve interface.

Besides the viscosity term, $\tau$ also exits in the error terms of the SC and the EDM force schemes. For the SC force scheme, the last term ($\delta_t \partial_i \left[ \delta_t \left( \tau - \frac{1}{2} \right)^2 \frac{F_i F_j}{\rho} \right]$) of the error terms $R_{v,sc}$ (Eq. 26) is not necessarily zero under any circumstances. Hence the dependence on $\tau$ always exist for the SC force scheme. For the EDM force model, dependence on $\tau$ through the error term $R_{v,EDM}$ comes from the $\partial_i \upsilon \left( \delta_t - \frac{1}{2} \right) \left( \partial_i F_j + \partial_j F_i \right)$. Similar to the viscosity term, $\partial_i \left( \partial_i F_j + \partial_j F_i \right)$ is zero for straight interfaces, so the error term will be independent of $\tau$ in this case. However, for curved interfaces $\partial_i \left( \partial_i F_j + \partial_j F_i \right)$ is not necessarily zero, leading to expected dependence of density distribution on the value of $\tau$. To verify these observations, numerical results obtained using these force schemes are compared, and presented in the next section. Suggestions are then made to reduce the spurious velocities and the effect of $\tau$ on the numerical results.

## 4. Numerical simulations

In this section, we numerically compare the performance of the three force schemes for their accuracy and to determine how strongly the results depend on $\tau$. A different force scheme is suggested to improved Kupershtokh et al.'s model. Numerical results of the modified scheme are compared with Li et al.'s model. Finally, some suggestions are proposed to improve the accuracy of the modified Kupershtokh et al.'s model and to reduce the $\tau$'s effect and the spurious velocities.

**4.1 Incorporation of EOS**

In order to numerically investigate the performance of the interparticle force methods and different force schemes, the C-S EOS is applied in the present work, which can be written as [19]

$$p = \rho RT \frac{1 + b\rho/4 + (b\rho/4)^2 - (b\rho/4)^3}{(1 - b\rho/4)^3} - a\rho^2, \quad (29)$$

where $a = 0.4963 R^2 T_c^2 / p_c$, $b = 0.18727 RT_c / p_c$. The parameters are chosen as $a = 1$, $b = 4$, $R = 1$ [19].

It has been shown [23-25] that the scale of the EOS can influence the stability, and impact the interface width of the pseudo-potential model. Hence, a simple EOS scale adjustment [25] is applied in the present work. The C-S EOS is rewritten by introducing a scaling parameter $K$ as [25]:

$$p = K\left(\rho RT \frac{1 + b\rho/4 + (b\rho/4)^2 - (b\rho/4)^3}{(1 - b\rho/4)^3} - a\rho^2\right). \quad (30)$$

For $0 < K < 1$, the stability of this model can improve significantly [25], and the width of the interface is also increased. It should be pointed out that the Maxwell construction density solution (Eq.19) will not be changed for $K \neq 1$, however, the mechanical solution (Eq. 18) will be different.

**4.2 Comparison of different force schemes**

Here we test the three different force schemes with Kupershtokh et al.'s pseudo-potential model, and numerically compare the performance of these schemes for different $\tau$ and temperature. Based on the simulation results, the best force scheme is then adopted for improvement of Kupershtokh et al's model.

**4.2.1 Influence of relaxation time**

To assess the influence of relaxation time $\tau$ on different force schemes, we simulated the phase coexistence with different value of $\tau$. To avoid the influence of surface tension, a straight interface is simulated. Parameters used in this simulation of EDM and Shan and Chen's force schemes are given in Table 2. To compare these methods on different aspects and to maintain the stability, the parameters of Guo et al.'s method may be different for other cases. Since the gas phase is more compressible, the influence of $\tau$ will be more pronounced in gas phase, hence we only show the simulation results of gas density.

Table 2. Parameters used in EDM for the straight interface simulation

|   | EDM or SC |
|---|---|
| A | -0.26 |
| $\varepsilon$ | 0.52 |
| K | 1 |

The simulation results of gas density as a function of $\tau$ are shown in Fig. 2. For EDM and Guo's force schemes, the simulation gas densities do not change with the relaxation times, but for Shan-Chen's force scheme, the gas density increases obviously with the relaxation time. Moreover, the simulation results are identical for the SC and the EDM force schemes when $\tau = \delta_t = 1$. These results agree with the Chapman-Enskog expansion and Kupershtokh et al.'s work [23]. It also should be mentioned that the stability of SC force scheme decreases significantly when the relaxation time is relatively small.

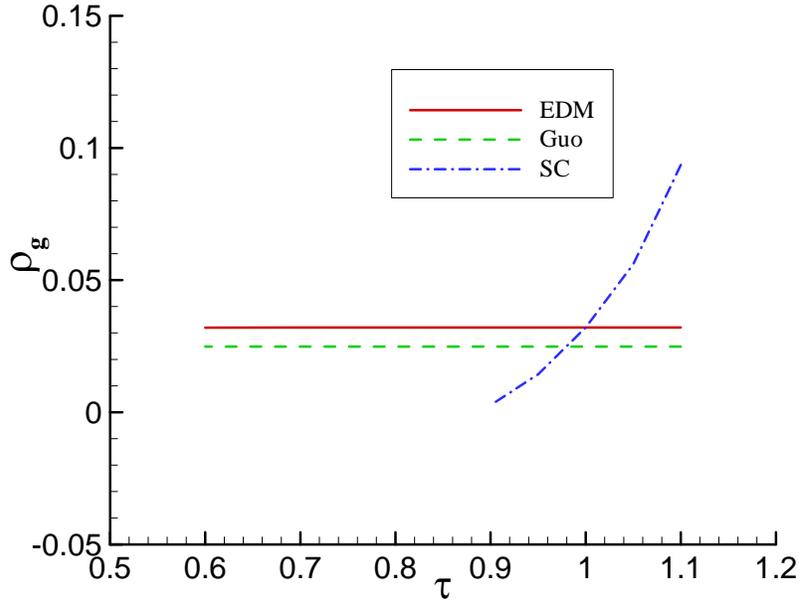

Fig. 2. Gas densities for different force schemes.

The EDM force is known to be influenced by $\tau$ when the density ratio is large [21]. To study the influence of $\tau$, the density ratio for EDM and Guo et al.'s force model under different temperatures is numerically analyzed. The parameters are still given by table 2, except, to maintain the stability of the model, $A$ is chosen as -0.5 for Guo et al.'s force scheme. Figures 3 and 4 show the density ratios as a function of reduced temperature for two different relaxation times $\tau$ for EDM and Guo et al.'s schemes, respectively. It can be seen that the simulated density ratios are exactly the same for $\tau = 0.7$ and $\tau = 1.5$. It means that the simulated densities of both EDM and Guo et al.'s scheme are not influenced by $\tau$ when the interface is straight. These results agree with the analyses in Section. 3.

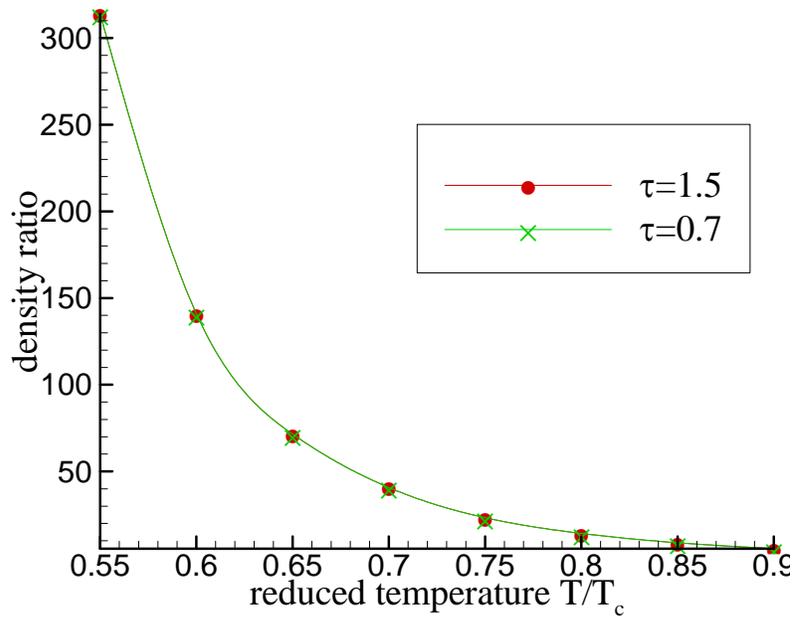

Fig. 3. Density ratio as a function of reduced temperature for two different values of relaxation time $\tau$ for EDM force scheme.

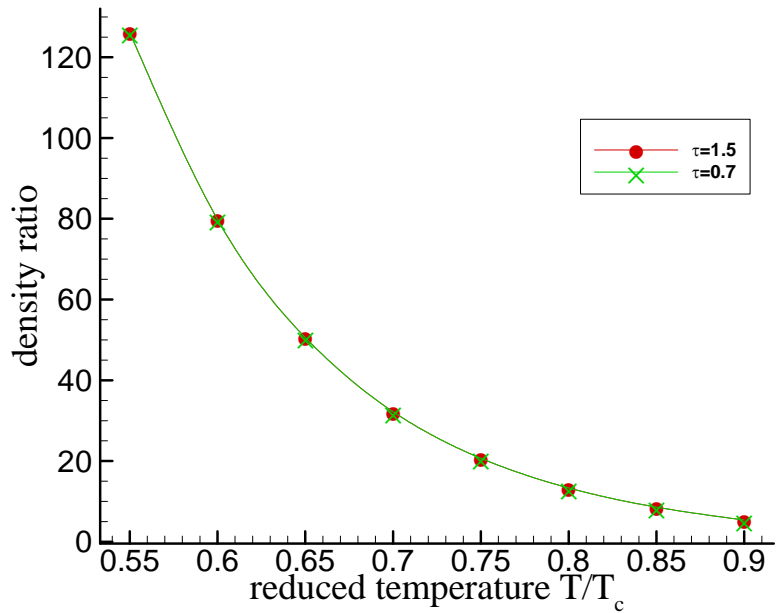

Fig. 4. Density ratio as a function of reduced temperature for two different values of relaxation time $\tau$ for Guo et al.'s force scheme.

To study the case of curved interface, a static bubble with radius of 23 l.u. (lattice units) was numerical simulated in a $100 \times 100$ lattice domain. The parameter for the EDM force scheme are: $A = -0.13$, $K = 0.5$, $T = 0.6Tc$. The parameters for Guo et al.'s scheme, to match the density ratio with the results of EDM force scheme, are chosen as: $A = -0.45$, $K = 0.5$, $T = 0.6Tc$. Results for two different values of $\tau$ are shown in Table. 3 ($Us$ represents the spurious velocity). It can be seen from Table. 3 that the surface tensions and maximum spurious velocities of these two schemes are both influenced by $\tau$. Consequently, the density distributions are influenced by $\tau$. Meanwhile, the influence of $\tau$ in Guo et al.'s scheme is in general smaller than in EDM force scheme. As discussed in Section 3.1, the influence of $\tau$ on Guo et al.'s scheme is caused by spurious velocities, while the influence of $\tau$ on EDM is coursed by spurious velocities and the effect of $\tau$ on the error terms ($R_{v, EDM}$). This might explain why the influence of $\tau$ on Guo et al.'s scheme is smaller than on EDM force scheme.

Table. 3 Bubble simulation for two different value of $\tau$

| Force scheme | $\tau$ | $\rho_g$ | $\rho_l$ | $\sigma$ | $Us_{max}$ |
|---|---|---|---|---|---|
| Guo et al. | 1.5 | 0.002832 | 0.40403 | 0.01059 | 0.00874221 |
| Guo et al. | 0.7 | 0.003172 | 0.40403 | 0.010368 | 0.0203277 |
| EDM | 1.5 | 0.003406 | 0.403652 | 0.01241 | 0.0110391 |
| EDM | 0.7 | 0.002366 | 0.403605 | 0.013635 | 0.00495204 |

### 4.2.2 Comparison of solutions for different force schemes with mechanical solutions

In this section we compared the results obtained using the EDM and the Guo et al.'s force schemes with the mechanical solutions (Eq. 18). The parameters were chosen as follows: $\varepsilon = 0.52$, $A = -\varepsilon/2 = -0.26$, $K = 1$. Since the mechanical analysis is based on a straight interface, only straight interface cases were considered here.

The simulation results of gas densities for different temperatures for these two force schemes are shown in Fig. 5. It can be seen that the results of Guo et al.'s force scheme matches the mechanical solution much better than the EDM force scheme's results. These results are consistent with our analysis in Sec. 3 as well as with Li et al. [3].

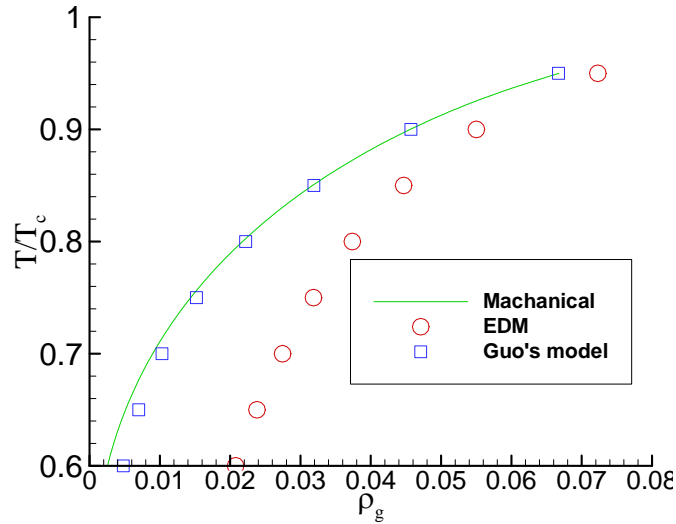

Fig. 5 Gas densities as a function of temperature for these two force schemes compared with the mechanical solution

Based on the above simulation results, we can see that although $\tau$ has no influence on both EDM and Guo et al.'s force schemes when the interface is straight, the simulation results vary with $\tau$ when the interface is curved, and the influence is more pronounced on the EDM force scheme. In addition, solutions obtained using the Guo et al.'s force scheme agree with the mechanical solutions much better than those obtained using the EDM scheme. Based on these results, Guo et al.'s force scheme is selected for application in the following simulations.

**4.3 Comparison with Li et al.'s model**

To improve the stability of the pseudo-potential model, Li et al. [3] proposed a model to adjust $\varepsilon$ by modifying Guo et al.'s force scheme. In this model, the interparticle interaction force is calculated by Eq. (4), and the force term in LBE is given by [3]

$$\boldsymbol{F}_\alpha = \omega_\alpha \left(1 - \frac{1}{2\tau}\right) \left[ \frac{F_i e_{\alpha i}}{c_s^2} + \frac{(F_i v'_j + v'_j F_i)(e_{\alpha i} e_{\alpha j} - c_s^2 \delta_{ij})}{2 c_s^4} \right], \quad (31)$$

where $v' = v + \sigma \boldsymbol{F}/(\upsilon \psi^2)$, and $\sigma$ is a constant which determines the value of $\varepsilon$. The relationship between $\varepsilon$ and $\sigma$ is given by

$$\varepsilon = -2(\alpha + 24\delta t G \sigma)/\beta. \quad (32)$$

It should be noted that by choosing this model, the simulation results will now be dependent on *G*. In the present work, *G* was chosen as -1.

The principle of Li et al.'s model is the same as that of Kupershtokh et al.'s model, which improves the stability by adjusting $\varepsilon$ in the mechanical solution (Eq. 18). However, their performances may be different at large density ratios since the pressure tensor error terms of Li et al.'s model (Eq. (14)) is different from the pressure tensor error terms of Kupershtokh et al.'s model (Eq. (16)). Different characteristics of these two models are compared here.

### 4.3.1 Thermodynamic consistency and mechanical solution

To satisfy the thermodynamic consistency, the mechanical solution should match the Maxwell construction. To satisfy this requirement, Shragaglia and Shan [30] proposed an effective density $\psi$ as

$$\psi(\rho) = \begin{cases} \exp(-1/\rho), & \varepsilon = 0 \\ \left(\dfrac{\rho}{1+\rho}\right)^{1/\varepsilon}, & \varepsilon \neq 0 \end{cases}. \tag{33}$$

This form of $\psi(\rho)$ makes Eq. (18) and Eq. (19) to be identical. However, this choice limits the choice of the EOS in the pseudo-potential model. To apply the model to different EOSs, Li et al. [3] pointed out that by adjusting $\varepsilon$ in the mechanical solution, the thermodynamic consistency can be approximately satisfied.

To compare the accuracy of the modified Kupershtokh et al.'s model and Li et al.'s model, we first simulated the gas densities for the straight interface case with the same $\varepsilon$, and compared the results with the mechanical solution and Maxwell construction solution. In these simulations, $\varepsilon = 1.68$ [3], accordingly, $A = 0.84$, $\sigma = 0.105$, and $K = 1$.

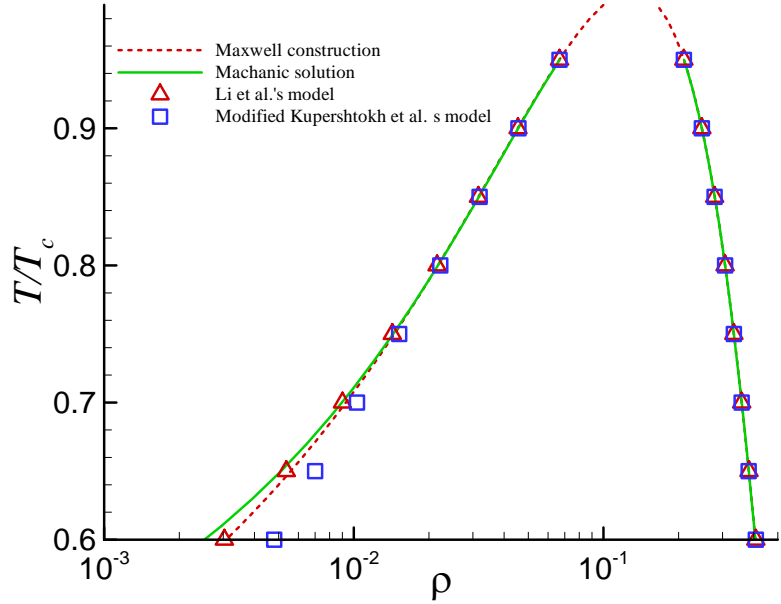

Fig. 6 Comparison of coexistence curves obtained using Kupershtokh et al.'s and Li et al.'s model.

Figure. 6 shows the coexistence curves obtained using these two models for different temperatures. It can be seen that these results agree well with each other for the liquid branch (right side) and for the relatively high temperature values of the gas branch (left side). However, at low temperature values (T/Tc < 0.7), the difference between the mechanical solution and the simulation results become noticeable for the gas branch, and the results of Li et al.'s model are closer to the mechanical results. The reason behind the discrepancy at low temperature values may be the higher order error terms of the pressure tensor in Li et al.'s model (Eq. 14) are different from the corresponding terms in the modified Kupershtokh et al.' model (Eq. 16), and these error terms become large when the temperature is too low since the density changes rapidly over the thin interface.

**4. 3. 2  $\tau$'s effects**

The results above show a deviation in results obtained using between these two models when the temperatures are low. Hence, we further assess the performance of these two models for low temperature values (T < 0.7Tc) and for curved interfaces. Specifically, we study the effects of $\tau$ on the performance of these two models for $\tau = 0.7, 0.8, 1.0$ and $1.5$, The parameters chosen are the same as those in the previous section except initially, a single bubble is placed at the center of simulation domain of 100×100 lattices. The radius of the bubble is 23 l.u. and the temperature is equal to 0.6 $T_c$ in the simulation.

Simulation results are shown in Table 4 and Table 5 (Li et al.'s model becomes unstable for $\tau$ is equal to 0.7). It can be seen that the influence of $\tau$ is more pronounced on gas densities for Li et al's model, and the $\tau$'s effect on density ratios is also larger. However, the largest spurious velocity obtained by these models vary for different $\tau$: when $\tau < 1$, the maximum spurious velocity in Kupershtokh et al.'s model is larger than that in Li et al.'s model; when $\tau > 1$, the largest spurious velocity in Kupershtokh et al.'s model is smaller than in Li et al.'s model. Hence, conclusion can be made that although Kupershtokh et al.'s model shows lower accuracy compared to Li et al.'s model for straight interface case when the temperature is low, it is more stable for curved interfaces.

Table 4. Influence of $\tau$ on Kupershtokh et al.'s model

| $\tau$ | $\rho_l$ | $\rho_g$ | $Us_{max}$ |
|---|---|---|---|
| 0.7 | 0.407189 | 0.00515232 | 0.117 |
| 0.8 | 0.406585 | 0.0047108 | 0.0427 |
| 1 | 0.406599 | 0.00429998 | 0.0087 |
| 1.5 | 0.406635 | 0.00430357 | 0.0211 |

Table 5. Influence of $\tau$ on Li et al.'s model

| $\tau$ | $\rho_l$ | $\rho_g$ | $Us_{max}$ |
|---|---|---|---|
| 0.8 | 0.406113 | 0.00282034 | 0.012 |
| 1 | 0.406128 | 0.00337611 | 0.028 |
| 1.5 | 0.406071 | 0.00485067 | 0.0396 |

Over all, Li et al.'s model and the modified Kupershtokh et al's model have the same theoretical base, and their simulation results agree well with each other when the temperature is relatively high. However, the performances of these models are still different when the temperature is low.

**4.4 Possible approach to improve the accuracy and to reduce the $\tau$'s effect for low temperatures**

Through the accuracy of Kupershtokh et al.'s model can be improved by applying Guo et al.'s force scheme, the simulation results still do not match the mechanical solution perfectly, and the $\tau$'s influence cannot be eliminated for the low temperatures. It is difficult to completely eliminate the drawbacks due to the mechanical nature of the pseudo-potential model, however, the shortcomings can be further reduced. Here we proposed two possible methods to improve the accuracy and to reduce the $\tau$'s effect on the model.

**4.4.1 Scale of the EOS**

Since the difference between the simulation results and the construction solution becomes noticeable for the thin interface corresponding to low temperatures, the error can be reduced by enlarging the interface width by rescaling the EOS. This can be achieved by changing the value of the parameter K in EOS. (It should be noted that the $\varepsilon$ should also be changed to approximately match the Maxwell construction). Hence here we compared the results for two values of $K$ (1 and 0.1) with mechanical solution. The temperature is $T/Tc = 0.6$, corresponding interface widths are about 3 and 6 lattice units, and the values of $\varepsilon$ are 1.68 and 2 for $K$ equal to 1 and 0.1, respectively. The simulation results are presented in Table. 6. It can be seen that when $K = 0.1$, the difference between simulation results and the mechanical solution are much reduced compared with the case for $K = 1$.

Table. 6 Comparison of gas densities for different $K$ values

|  | Maxwell construction | Mechanical solution ($K=1$, $\varepsilon=1.68$) | Simulation results ($K=1$, $\varepsilon=1.68$) | Mechanical solution ($K=0.1$, $\varepsilon=2$) | Simulation results ($K=0.1$, $\varepsilon=2$) |
|---|---|---|---|---|---|
| $\rho_g$ | 0.00308 | 0.00245 | 0.0048 | 0.00310 | 0.00345 |

**4.4.2 Multiple relaxation time (MRT) model**

Considering the $\tau$'s value is only related to the viscosity of fluid, it is possible to reduce the $\tau$'s effects by applying the MRT model [32, 33] since only two of its nine relaxation time parameters need to be changed to adjust the viscosity. To transfer Guo's force scheme into MRT formulation, the collision step of LBE in moment space can be written as [34-36]:

$$\bm{m}^* = \bm{m} - \Lambda(\bm{m} - \bm{m}^{eq}) + \delta_t\left(\bm{I} - \frac{\Lambda}{2}\right)\bm{S}, \tag{34}$$

where $\bm{m} = \bm{M}\bm{f}$, $\bm{m}^{eq} = \bm{M}\bm{f}^{eq}$, $\bm{S} = \bm{M}\bm{F}/\left(1 - \frac{1}{2\tau}\right)$, and $\bm{M}$ is the transformation matrix. The diagonal matrix $\Lambda$ is given by

$$\Lambda = diag(\tau_\rho^{-1}, \tau_e^{-1}, \tau_\varsigma^{-1}, \tau_j^{-1}, \tau_q^{-1}, \tau_j^{-1}, \tau_q^{-1}, \tau_\upsilon^{-1}, \tau_\upsilon^{-1}), \tag{35}$$

where only $\tau_\upsilon$ is related to viscous and it should equal to $\tau$ in the corresponding LBGK equation. The streaming process is given by

$$f_\alpha(\bm{x} + \bm{e}_\alpha \Delta t, t + \Delta t) = f^*(\bm{x}, t), \tag{36}$$

where $\bm{f}^* = \bm{M}^{-1}\bm{m}^*$.

To assess the $\tau_\upsilon$'s effects on the MRT model, a static bubble is simulated with different values of $\tau_\upsilon$. To compare the results with LBGK model, the same parameters are chosen as used in the simulation of the results presented in Table. 4. It can be seen from Table 7 that the effect of $\tau_\upsilon$ is significantly reduced compared to LBGK, and the largest spurious velocities are also relatively much smaller in magnitude.

Table 7. Static bubble simulation of MRT model

| $\tau_\upsilon$ | $\rho_l$ | $\rho_g$ | $Us_{max}$ |
|---|---|---|---|
| 0.7 | 0.404967 | 0.00220167 | 0.0069 |
| 1 | 0.404963 | 0.00213255 | 0.0032 |
| 1.5 | 0.404972 | 0.00205247 | 0.0048 |

**5 Summary and conclusions**

In this paper, we studied Kupershtokh et al.'s single component pseudo-potential lattice Boltzmann models in detail. Three primary force schemes were theoretically analyzed and their numerical results compared. Based on the results, Guo et al.'s force scheme was adopted to improve the accuracy of the Kupershtokh et al.'s model. Numerical comparisons have also be carried out between the modified model and Li et al.'s model. The simulation results show that although the high precision force scheme can eliminate the error terms in the corresponding momentum equation, the numerical errors and effect of $\tau$ are still noticeable when the temperatures are relatively low, especially when interfaces are curved. These errors are possibly because the viscosity terms in the momentum equation are dependent on spurious velocities. Besides, the high order error terms in the pressure tensor also make the simulation results different from the mechanical solutions. These error terms lead to different performances between the modified Kupershtokh et al.'s model and Li et al.'s model. To improve the accuracy of the modified model and to reduce the effects of $\tau$, we proposed two approaches: increasing interface width, and applying the MRT model. Simulation results show that the performance of the model can be further improved by applying these two methods.

**Acknowledgements**

This work is sponsored by National Natural Science Foundation of China (51076172) and Science and Technology Innovation Foundation of Chongqing University (CDJXS11142232). The authors thank Dr. Q. Li for helpful discussions.

# References:

[1]. Kupershtokh, A.L., D.A. Medvedev and D.I. Karpov, On equations of state in a lattice Boltzmann method. Computers and Mathematics with Applications, 2009. 58(5): p. 965 - 974.

[2]. Guo, Z., C. Zheng and B. Shi, Discrete lattice effects on the forcing term in the lattice Boltzmann method. Physical Review E, 2002. 65(4): p. 046308.

[3]. Li, Q., K.H. Luo and X.J. Li, Forcing scheme in pseudopotential lattice Boltzmann model for multiphase flows. Physical Review E, 2012. 86(1): p. 016709.

[4]. Guy, R., McNamara and G. Zanetti, Use of the Boltzmann Equation to Simulate Lattice-Gas Automata. Physical Review Letters, 1988. 61(20): p. 2332 - 2335.

[5]. Gunstensen, A.K., et al., Lattice Boltzmann model of immiscible fluids. Physical Review A, 1991. 43(8): p. 4320 - 4327.

[6]. Shan, X. and H. Chen, Lattice Boltzmann model for simulating flows with multiple phases and components. Physical Review E, 1993. 47(3): p. 1815-1819.

[7]. Swift, M.R., et al., Lattice Boltzmann simulations of liquid-gas and binary fluid systems. Physical Review E, 1996. 54(5): p. 5041-5052.

[8]. Swift, M.R., W.R. Osborn and J.M. Yeomans, Lattice Boltzmann Simulation of Nonideal Fluids. Physical Review Letters, 1995. 75(5): p. 830 - 833.

[9]. He, X., X. Shan and G.D. Doolen, Discrete Boltzmann equation model for nonideal gases. Physical Review E, 1998. 57(1): p. R13-R16.

[10]. He, X. and G.D. Doolen, Thermodynamic Foundations of Kinetic Theory and Lattice Boltzmann Models for Multiphase Flows. Journal of Statistical Physics, 2002. 107(1): p. 309 - 328.

[11]. Luo, L., Unified Theory of Lattice Boltzmann Models for Nonideal Gases. Physical Review Letters, 1998. 81(8): p. 1618 - 1621.

[12]. Tölke, J., Lattice Boltzmann simulations of binary fluid flow through porous media. Philosophical Transactions of the Royal Society of London. Series A: Mathematical, Physical and Engineering Sciences, 2002. 360(1792): p. 535-545.

[13]. Alexander, F.J., S. Chen and D.W. Grunau, Hydrodynamic spinodal decomposition: Growth kinetics and scaling functions. arXiv preprint comp-gas/9304005, 1993.

[14]. Guo, Z. and C. Shu, Lattice Boltzmann Method and Its Applications in Engineering. 2013: World Scientific.

[15]. Lee, T. and C. Lin, A stable discretization of the lattice Boltzmann equation for simulation of incompressible two-phase flows at high density ratio. Journal of Computational Physics, 2005. 206(1): p. 16 - 47.

[16]. Chao, J., et al., A filter-based, mass-conserving lattice Boltzmann method for immiscible multiphase flows. International Journal for Numerical Methods in Fluids, 2011. 66(5): p. 622-647.

[17]. Zheng, H.W., C. Shu and Y.T. Chew, A lattice Boltzmann model for multiphase flows with large density ratio. Journal of Computational Physics, 2006. 218(1): p. 353 - 371.

[18]. Inamuro, T., et al., A lattice Boltzmann method for incompressible two-phase flows with large density differences. Journal of Computational Physics, 2004. 198(2): p. 628 - 644.

[19]. Yuan, P. and L. Schaefer, Equations of state in a lattice Boltzmann model. Physics of Fluids, 2006. 18(4): p. 42101.

[20]. Shan, X. and H. Chen, Simulation of nonideal gases and liquid-gas phase transitions by the lattice Boltzmann equation. Physical Review E, 1994. 49(4): p. 2941-2948.

[21]. Huang, H., M. Krafczyk and X. Lu, Forcing term in single-phase and Shan-Chen-type multiphase lattice Boltzmann models. Physical review. E, Statistical, nonlinear, and soft matter physics, 2011. 84(4 Pt 2): p. 046710.

[22]. Sbragaglia, M., et al., Generalized lattice Boltzmann method with multirange pseudopotential. Physical review. E, Statistical, nonlinear, and soft matter physics, 2007. 75(2 Pt 2): p. 026702.

[23]. Kupershtokh, A.L., A lattice Boltzmann equation method for real fluids with the equation of state known in tabular form only in regions of liquid and vapor phases. Computers and Mathematics with Applications, 2011. 61(12): p. 3537 - 3548.

[24]. Kupershtokh, A.L., Criterion of numerical instability of liquid state in LBE simulations. Computers and Mathematics with Applications, 2010. 59(7): p. 2236 - 2245.

[25]. Hu, A., et al., On equations of state in pseudo-potential multiphase lattice Boltzmann model with large density ratio. International Journal of Heat and Mass Transfer, 2013. 67: p. 159 - 163.

[26]. Kupershtokh, A.L., et al., Stochastic models of partial discharge activity in solid and liquid dielectrics. IET Science, Measurement&Technology, 2007. 1(6): p. 303 - 311.

[27]. Qian, Y.H., et al., Lattice BGK Models for Navier-Stokes Equation. EPL (Europhysics Letters), 1992. 17(6): p. 479 - 484.

[28]. Zhang, R. and H. Chen, Lattice Boltzmann method for simulations of liquid-vapor thermal flows. Physical review. E, Statistical, nonlinear, and soft matter physics, 2003. 67(6): p. 066711.

[29]. Shan, X., Pressure tensor calculation in a class of nonideal gas lattice Boltzmann models. Physical Review E, 2008. 77(6): p. 066702.

[30]. Shan, X. and M. Sbragaglia, Consistent pseudopotential interactions in lattice Boltzmann models. Physical Review E, 2011. 84(3): p. 036703.

[31]. Guo, Z. and C. Zheng, Theory and Applications of Lattice Boltzmann Method. 2009, Beijing: Science Press.

[32]. Luo, L. and P. Lallemand, Theory of the lattice Boltzmann method: Dispersion, dissipation, isotropy, Galilean invariance, and stability. Physical Review E, 2000. 61(6): p. 6546-6562.

[33]. D'Humieres, D., Generalized lattice-Boltzmann equations. Rarefied gas dynamics- Theory and simulations, 1994: p. 450-458.

[34]. Abraham, J. and M.E. McCracken, Multiple-relaxation-time lattice-Boltzmann model for multiphase flow. Physical Review E, 2005. 71(3): p. 036701.

[35]. Fan, L. and Z. Yu, Multirelaxation-time interaction-potential-based lattice Boltzmann model for two-phase flow. Physical Review E, 2010. 82(4): p. 046708.

[36]. Li, Q., K.H. Luo and X.J. Li, Lattice Boltzmann modeling of multiphase flows at large density ratio with an improved pseudopotential model. Physical Review E, 2013. 87(5): p. 053301.